\documentstyle[11pt,aaspp4,flushrt]{article}
\def\etal{{et al.}\ }
\def\hst{{\it HST\/ }}
\def\hal{H$\alpha$}
\def\hbeta{H$\beta$}
\def\nion{$N_{ion}$}
\def\nhal{$N_{\rm H\alpha}$}

\begin{document}

\title{A Search for Ultraviolet Emission from LINERs}

\author{Aaron J. Barth}
\affil{Department of Astronomy, University of California, Berkeley CA
94720-3411}
\author{Luis C. Ho}
\affil{Harvard-Smithsonian Center for Astrophysics, 60 Garden Street,
Cambridge, MA 02138}
\author{Alexei V. Filippenko}
\affil{Department of Astronomy, University of California, Berkeley CA
94720-3411}
\author{Wallace L. W. Sargent}
\affil{Palomar Observatory, 105-24 Caltech, Pasadena, CA 91125}

\begin{abstract}

We have obtained {\it Hubble Space Telescope} Wide Field and Planetary
Camera 2 ultraviolet (UV) 2200 \AA\ and optical $V$-band images of 20
low-luminosity active galactic nuclei, most of which are
spectroscopically classified as LINERs, in order to search for a
possible photoionizing continuum.  Six (30\%) of the galaxies are
detected in the UV.  Two of the detected galaxies (NGC 3642 and NGC
4203) have compact, unresolved nuclear UV sources, while the remaining
four UV sources (in NGC 4569, NGC 5005, NGC 6500, and NGC 7743) are
spatially extended.  Combining our sample with that of Maoz \etal
(1995), we find that the probability of detection of a nuclear UV
source is greatest for galaxies having low internal reddening and low
inclination, and we conclude that dust obscuration is the dominant
factor determining whether or not a UV source is detected.  Large
emission-line equivalent widths and the presence of broad-line
emission also increase the likelihood of detection of nuclear UV
emission.  Our results suggest that the majority of LINERs harbor
obscured nuclear UV sources, which may be either accretion-powered
active nuclei or young star clusters.  Under the assumption that the
compact UV sources in NGC 3642 and NGC 4203 have nonstellar spectra of
slope $f_\nu \propto \nu^{-1}$ extending into the extreme ultraviolet,
the extrapolated ionizing fluxes are sufficiently strong to
photoionize the narrow-line regions of these objects. The $V$-band
images of many galaxies in our sample reveal remarkably strong dust
lanes which may be responsible for obscuring some UV sources.

\end{abstract}


To appear in {\it The Astrophysical Journal.}

\section{Introduction}

Emission-line galactic nuclei are generally classified into three
major categories: star-forming or \ion{H}{2} nuclei, in which young,
massive stars provide the photoionizing continuum; Seyfert nuclei,
which are generally understood to be the result of accretion onto a
supermassive black hole; and low-ionization nuclear emission-line
regions, or LINERs, which were first defined as a class by Heckman
(1980).  Despite much observational and theoretical study,
understanding the excitation mechanism of LINERs has proved to be an
elusive goal.  Optical spectroscopy alone does not provide sufficient
constraints to determine unambiguously the nature of LINERs.  Models
based on shock excitation (\cite{ko76}; \cite{fos78}; Dopita \&
Sutherland 1995, 1996), photoionization by hot stars (\cite{tm85};
\cite{ft92}; \cite{shi92}), and photoionization by a nonstellar
power-law continuum (\cite{fn83}; \cite{hs83}; Ho, Filippenko, \&
Sargent 1993) have all had reasonable success at reproducing the
optical emission-line ratios of LINERs.  There are compelling reasons
to believe that all of these mechanisms may play a role, but it is not
known which mechanism dominates in the majority of LINERs.

At the heart of the subject is the fundamental question of whether
LINERs are powered by accretion onto a black hole, or by massive
stars, or by some other process.  Ultraviolet (UV) observations can
shed light on this issue in several ways.  If a cluster of young stars
is present, then it should be possible to detect UV spectroscopic
signatures of massive stars such as P Cygni line profiles.  On the
other hand, an accretion-powered object, if not obscured by dust,
should manifest itself as a spatially unresolved source of a
featureless UV continuum, possibly accompanied by the broad emission
lines that are the signature of Seyfert and quasar activity.

A UV imaging survey using the pre-refurbishment {\it Hubble Space
Telescope (HST)} Faint Object Camera (FOC) and F220W filter
(\cite{mao95}) has shown that about 20\% of nearby LINERs harbor a
source of UV emission with a diameter of at most several parsecs.
Without spectroscopy, compact young star clusters cannot be ruled out
as the sources of the emission, but this result is highly suggestive
of the existence of genuine AGN-like featureless continua in at least
some LINERs.  Assuming an $f_{\nu} \propto \nu^{-1}$ continuum, Maoz
\etal find that the inferred ionizing luminosity is sufficient to
power the observed emission lines.  Maoz \etal proposed three possible
explanations for the low UV detection rate.  The majority of LINERs in
which UV emission is not detected could be shock excited rather than
photoionized; they could contain UV sources which are obscured by
dust; or, the central photoionizing source could be ``turned off'' the
majority of the time.  Eracleous, Livio, \& Binette (1995) have
explored this third possibility by proposing a model in which the
occasional disruption of a star by the central massive black hole
leads to the formation of an accretion disk lasting a few decades,
providing the energy source for the LINER, while the emission lines
decay on 100-year timescales.  Of course, these possible explanations
are not mutually exclusive, and they all may play a role in
determining the observed UV properties of LINERs.

In order to further constrain the excitation mechanism of LINERs, and
to discover more candidates for UV spectroscopy, we have carried out
an \hst\ UV and optical imaging survey using the Wide Field and
Planetary Camera 2 (WFPC2).  In agreement with the results of the Maoz
\etal survey, we find that the majority of the nuclei in our sample
are completely dark in the UV.  Among the objects showing UV emission,
we find that two have unresolved nuclear sources which are
sufficiently bright to account for the emission-line fluxes by
photoionization.  The remaining four UV-detected nuclei are spatially
extended sources.  By combining our dataset with that of Maoz et al.,
we are able to assemble a sample of galaxies large enough to study the
relationships between the UV detection rate and overall host galaxy
properties such as inclination, reddening, and emission-line ratios.

\section{Sample Selection and Observations}

The galaxy sample consists of 20 LINERs, low-luminosity Seyfert 2
nuclei, and LINER/\ion{H}{2} ``transition'' objects selected from the
Palomar Observatory Dwarf Seyfert Survey (\cite{fs85}; \cite{hfs95}).
Objects were selected on the basis of having weak or undetected broad
\hal\ emission; the detailed profile-fitting analysis carried out by
Ho \etal (1997b) has subsequently detected broad \hal\ emission in
five of these galaxies.  Based on the classifications of Ho,
Filippenko, \& Sargent (1997a), the sample contains 14 LINERs, 3
Seyfert 2s, and 3 LINER/\ion{H}{2} transition objects.  The galaxies
span a range in Hubble types from S0 to Sc, and the median redshift of
the sample is $cz=1070$ km s$^{-1}$.  Five galaxies in this sample
were also observed as part of the FOC UV snapshot survey of Maoz \etal
(1995).  Table 1 lists the basic galaxy parameters and the WFPC2
exposure times.

The galaxies were observed between May 1994 and March 1995, through
the F218W filter (mean wavelength 2190 \AA, effective width 390 \AA),
with the nucleus placed on the Planetary Camera CCD (pixel scale
0\farcs046, field of view $35\arcsec\times35\arcsec$).  Exposures were
split into two separate integrations to facilitate cosmic-ray removal,
and total exposure times ranged from 1400 to 2200 s.  Brief (230-350
s) single exposures in the F547M ($V$ band) filter were also obtained
to verify pointing, to examine the optical morphology of the nuclei,
and to constrain the red leak contribution to the UV signal.  Bias
subtraction and flat-fielding were performed by the standard \hst\
data calibration pipeline.  Figure 1 shows the F218W images of the
UV-detected nuclei, and the F547M images are displayed in Figure 2.

For the UV images, cosmic ray removal and averaging of the two
sub-exposures were performed using the CRREJECT task in
IRAF\footnote{IRAF is distributed by the National Optical Astronomy
Observatories, which are operated by the Association of Universities
for Research in Astronomy, Inc., under cooperative agreement with the
National Science Foundation.}.  We removed cosmic-ray hits from the
F547M images by replacing abnormally high pixels with the average
count level found in neighboring pixels, using the IRAF COSMICRAYS
task, but many residuals and faint streaks remain in the images.

Distances to the galaxies, when needed, were obtained from the Nearby
Galaxies Catalog (\cite{tul88}), which assumes a Hubble constant of
$H_0 = 75$ km s$^{-1}$ Mpc$^{-1}$ and the Virgo infall model of Tully
\& Shaya (1984).  For NGC 6500, which is too distant to appear in the
catalog, we computed the distance using $H_0 = 75$ km s$^{-1}$
Mpc$^{-1}$.

\section{Results}

\subsection{UV Detection Rate}

After cosmic-ray cleaning, we blinked the UV and optical images of
each galaxy against each other to determine by eye whether UV emission
was detected at the nucleus.  Most images proved to be completely
blank in the UV; only four clearly showed nuclear UV emission, while
two others showed extremely weak features at the nucleus.  Off-nuclear
clusters of hot stars or foreground Galactic stars were visible in a
few others.  Because the UV and optical images of each galaxy were
obtained in a single telescope pointing, the position of the nucleus
on the PC detector should be the same in the UV and optical exposures.
We found this to be the case for each of the UV-detected galaxies.

We performed photometry on the F218W images of the UV-detected
galaxies using DAOPHOT (\cite{ste87}).  Total F218W counts were
measured in a circular aperture large enough to surround all the
detected flux from each nuclear source.  Count rates were then
converted to continuum fluxes at 2200 \AA\ using the relation that 1
count s$^{-1}$ is equivalent to $f_{\lambda}$(2200 \AA) $= 1.03 \times
10^{-15}$ erg cm$^{-2}$ s$^{-1}$ \AA$^{-1}$.  This conversion was
determined by performing synthetic photometry on a power-law continuum
with $f_{\nu} \propto \nu^{-1}$ using the F218W transmission curve in
the IRAF SYNPHOT package.  We found that varying the power-law slope
by $\pm1$ or substituting an O-type stellar continuum resulted in
variations of $<5\%$ in the counts-to-flux conversion.

Count rates were also corrected for the time-dependent WFPC2
throughput, which varies due to the buildup of contaminants in the
camera optics.  The mean throughput correction was 5\%, and no
observation required a correction of greater than 10\%.  Extinction
corrections were computed using the reddening estimates of Burstein \&
Heiles (1984) and the Galactic reddening curve of Cardelli, Clayton,
\& Mathis (1989).  The measured and extinction-corrected fluxes are
listed in Table 2.

Two of the UV-detected galaxies, NGC 3642 and NGC 4203, contain
nuclear sources that appear pointlike, or nearly so, but there is
clearly some faint extended UV emission surrounding the nucleus of NGC
3642.  Figure 3 shows the photometric growth curves for these two
sources and for an artificial point-spread function (PSF) created
using TINY TIM (version 3.0a; \cite{kr94}), normalized to the same
magnitude within a radius of $r=1$ pixel.  The profiles of NGC 3642
and 4203 are nearly identical out to $r=2$ pixels radius, and are only
marginally more extended than the artificial PSF.  We conclude that
the cores of these sources ($r\leq2$ pixels) are not resolved by {\it
HST.}  At larger radii ($r>3$ pixels), the profiles reveal the
presence of spatially extended emission, particularly in NGC 3642.  By
comparing the radial profiles of the UV sources with those of
artificially broadened PSFs, we estimate that the cores of these
sources, which dominate the light at $r\leq2$ pixels radius, have
intrinsic widths of $<0.5$ pixel full-width at half-maximum (FWHM).
The corresponding physical size limits are $<3.1$ pc and $<1.1$ pc
FWHM at the distances of NGC 3642 and NGC 4203 (27.5 and 9.7 Mpc,
respectively).

The other two easily detected galaxies, NGC 4569 and NGC 6500, are
clearly extended sources.  The UV-emitting region in NGC 6500 is a
diffuse patch without a bright central core.  In NGC 4569, the UV
source is centrally peaked but elongated along PA $\approx20\arcdeg$;
the photometric growth curve illustrated in Figure 3 shows that it is
more extended than the UV sources in NGC 3642 and NGC 4203.  Its FWHM
size is 13 $\times$ 9 pc$^2$ along its major and minor axes (at a
distance of 16.8 Mpc), while the diffuse source in NGC 6500 has a
diameter of $\sim90$ pc (at a distance of 40 Mpc).

The UV images of NGC 5005 and NGC 7743 show very faint excesses over
the background level at the positions of the optical nuclei.  These
features are not likely to be due solely to noise or cosmic rays, as
they are located precisely at the positions of the optical nuclei and
they appear to be diffuse sources, rather than residual cosmic-ray
spikes.  From our photometry, we estimate the significance of these
detections to be 6$\sigma$ for NGC 7743 and 4$\sigma$ for NGC 5005, as
measured in a $7\times7$ pixel detection aperture.  A diameter of 7
pixels for these sources corresponds to physical dimensions of 33 pc
and 38 pc, respectively, at distances of 21.3 Mpc and 24.4 Mpc for NGC
5005 and NGC 7743.

We determined upper limits to the F218W count rates of the undetected
galaxies by examining the noise level of the images and by artificial
star tests, assuming point-source profiles.  The two major sources of
noise in the UV images are readout noise and cosmic-ray residuals.
The mean background noise level measured in the UV images (after
cosmic-ray removal and averaging of sub-exposures) is 0.75 counts
pixel$^{-1}$, with a dispersion of 0.10 counts pixel$^{-1}$ among the
images.  As the noise properties of all the UV images are similar, we
assume for simplicity that all images of undetected galaxies have
$\sigma=0.75$ counts pixel$^{-1}$.  Then, the $3\sigma$ upper limit to
the flux of a UV point source in the undetected galaxies is 10 total
counts within a detection aperture of $r=2$ pixels.  We also
experimented with adding artificial stars with a range of magnitudes
to the images, and found that stars with 10 total counts were not
always found by eye, while stars with 20 or more total counts were
consistently detectable.  We therefore adopted 20 total counts as our
conservative upper limit to the flux of any nuclear UV point source in
the undetected galaxies.  This count limit was then converted to flux
limits in the same manner as described above, and the results are
given in Table 2.

One further concern is the red leak of the F218W filter, particularly
for such weak detections as NGC 5005 and NGC 7743.  The F218W
transmission curve extends to $\sim5000$ \AA, and for our purposes, we
will consider the red leak contribution to be the count rate measured
through the F218W filter due to photons with $\lambda>3000$ \AA.
Applying the measured transmission curve to a variety of stellar and
galactic spectra in SYNPHOT, we determined the expected red leak
contribution as a function of spectral type.  The fractional
contribution of red leak to the total F218W count rate ranges from
0.3\% for an A0 star to 58\% for an M6 star.  For a template
UV-optical spectrum of an S0 galaxy (provided by D. Calzetti), which
should be a reasonable match in overall color to most of the galaxies
in our sample, the red leak contribution to the F218W count rate is
2\% and the ratio of the F547M count rate to the F218W red leak count
rate is $1\times10^5$.  Most galaxies in our sample have F547M count
rates of $\lesssim3$ counts s$^{-1}$ in the peak pixel; the implied
red leak count rate of $\lesssim3\times10^{-5}$ counts s$^{-1}$ is far
too small to be detectable in exposure times typical of our sample.
For example, the brightest pixel in the nucleus of NGC 4203 has 4.2
counts s$^{-1}$ in F547M and 0.04 counts s$^{-1}$ in F218W; if our
estimates based on the S0 template spectrum are roughly correct, then
the red leak contribution to the F218W count rate should be
$\sim0.1\%$, or less than one count in the peak pixel.  Based on these
estimates, we are confident that the effects of red leak are
insignificant in our data.

The total UV detection rate for our sample is 30\%, while the FOC
survey of Maoz \etal found a detection rate of 20\% for LINERs and
LINER/\ion{H}{2} transition galaxies.  The greater detection rate in
the WFPC2 sample may be in part due to the longer integrations and
post-refurbishment optics.  A further advantage of the current survey
is that the optical images allow us to find the exact locations of the
nuclei in the UV images, hence we are able to identify very weak
nuclear sources such as those in NGC 5005 and NGC 7743 which might
otherwise have gone unnoticed.  If we had not detected these two very
faint sources, our detection rate would have been 20\%.  Combining the
WFPC2 and FOC samples, there are 24 objects classified as LINERs, and
six are detected in the UV, for an overall detection rate of 25\%.

\subsection{Host Galaxy Inclinations and  Obscuration}

The results of Maoz \etal (1995) and of this survey indicate that UV
emission is not detected in the majority of LINERs.  To what extent
might dust obscuration be responsible for this fact?  One
straightforward way to test the obscuration hypothesis is to compare
the host galaxy inclinations of the UV-detected and the UV-undetected
objects.  To improve the statistical significance of our results, we
have combined our WFPC2 sample with the sample of Maoz et al.,
selecting only spiral and S0 hosts for which an inclination can be
meaningfully measured.  The results are shown in Figure 4, for all
galaxies in the combined WFPC2 + FOC sample (excluding ellipticals),
and also for the subset of these galaxies which are classified as pure
LINERs.

It is immediately clear that the distributions of inclination are
different for the UV-detected and undetected galaxies.  In the entire
sample, the mean inclination for UV-detected galaxies is 40\arcdeg,
while for UV-undetected galaxies the mean inclination is 63\arcdeg.
For the LINERs-only subsample, the corresponding figures are
35\arcdeg\ and 62\arcdeg, respectively.  No galaxy with $i>65\arcdeg$
is detected in the UV, while 4 out of 6 LINERs with $i<45\arcdeg$ are
detected.  Applying the Kolmogorov-Smirnov (KS) test, we find a
probability of $P_{KS}=1.6\%$ that the distributions of inclinations
for UV-detected and undetected galaxies could be drawn from the same
parent population, or 1.8\% for the LINERs-only subsample.  Detection
of a UV source is thus significantly more likely in a low-inclination
host than in a highly inclined host, and obscuration by dust is the
most plausible explanation.  Since 2/3 of LINERs with $i<45\arcdeg$
are detected, it is reasonable to suppose that the true fraction of
LINERs having UV sources at their nuclei is $\gtrsim2/3$, but that
extinction hides the majority of the UV sources from our view.

To further test the obscuration hypothesis, we have compared the
Balmer decrements of the UV-detected and undetected LINERs, using the
line ratio data of Ho \etal (1997a), which have been corrected for
Galactic reddening using the Burstein \& Heiles (1984) correction
factors.  In order to remove possible sources of bias from the
comparison, we include only objects which are classified as LINERs by
Ho \etal (1997a).  The comparison is complicated by the fact that for
a few objects, the measured Balmer decrements are unphysical; that is,
\hal/\hbeta $< 2.8$, sometimes by a large margin.  Uncertainty in the
starlight subtraction procedure is the most likely cause of these
anomalous results, particularly for galaxies with extremely weak
\hbeta\ emission.  To minimize the influence of unphysical values on
the sample properties, we have computed median, rather than mean,
values of the Balmer decrements.  In the combined WFPC2 + FOC sample,
the UV-detected LINERs have a median \hal/\hbeta\ of 3.18 (for 6
objects), while the UV-undetected LINERs have a median \hal/\hbeta\ of
4.33 (18 objects).  Among the UV-detected LINERs, the largest Balmer
decrement is 3.51 (for NGC 404), while 70\% of the UV-undetected
LINERs have Balmer decrements greater than 3.5.  The distributions of
\hal/\hbeta\ for the UV-detected and undetected objects are shown in
Figure 5, and the KS test indicates that the distributions are
different at the 99\% confidence level ($P_{KS}=0.01$).

The median Balmer decrement of 4.33 for the UV-undetected LINERs
implies an extinction of $A_V=1.1$ mag, or 3.5 mag of extinction at
2200 \AA.  Such a large extinction would be more than sufficient to
completely hide from view UV sources as bright as those in NGC 3642
and NGC 4203.  We conclude that the UV-undetected LINERs are on
average more highly reddened by dust within the host galaxies than the
UV-detected LINERs, and that internal reddening is likely to be the
primary reason for the low UV detection rate.

\subsection{Equivalent Widths and Emission-Line Ratios}

Inspection of the optical spectra presented by Ho \etal (1995) reveals
a striking difference between the UV-detected and undetected LINERs:
in nearly all of the UV-detected LINERs, [\ion{O}{3}] is reasonably
strong and stands out clearly above the continuum, while in nearly all
of the UV-undetected objects, [\ion{O}{3}] and \hbeta\ appear to have
equivalent widths (EWs) so small that without starlight subtraction it
would be impossible to measure their strengths at all.  To quantify
this result, we show in Figure 6 the distributions of EW for \hbeta,
[\ion{O}{3}], and \hal\ for the UV-detected and undetected LINERs.
Median values of the EWs are listed in Table 3.  The KS test confirms
that the detected and undetected subsamples have different
distributions of EW, at least for [\ion{O}{3}] and \hbeta: $P_{KS}$ =
0.019, 0.039, and 0.14, respectively, for \hbeta, [\ion{O}{3}], and
\hal.  This result indicates that LINERs with greater emission-line
EWs are more likely to be detected in the UV.

We have also examined the distributions of other emission-line ratios
in order to search for other differences between the UV-detected and
undetected LINERs.  Using data from Ho \etal (1997a), we have compiled
the ratios [\ion{O}{3}] $\lambda5007$/\hbeta, [\ion{O}{1}]
$\lambda6300$/\hal, [\ion{N}{2}] $\lambda6583$/\hal, and [\ion{S}{2}]
$\lambda\lambda6716,6731$/\hal, corrected for both Galactic and
internal reddening.  To avoid biases due to a heterogeneous sample,
only objects classified as LINERs have been included.  Three objects
(NGC 5195, NGC 5322, and NGC 7814) were omitted because some lines
were not detected in their spectra.  Six UV-detected and 15
UV-undetected LINERs remain in the sample.  The distributions of line
ratios for the UV-detected and undetected objects are shown in Figure
5, and the median values of each distribution are listed in Table 3.
We note that the distribution of Hubble types does not differ
significantly between the UV-detected and undetected LINERs, so that
metallicity differences due to host galaxy type should not affect the
comparison.

To quantify whether there are any meaningful differences between the
emission-line ratios of the UV-detected and undetected subsamples, we
have applied the KS test to each emission-line ratio distribution; the
results for each line ratio are listed in Table 3.  The KS test
indicates that the [\ion{O}{1}]/\hal\ and [\ion{S}{2}]/\hal\ ratios do
not significantly differ between the UV-detected and undetected
LINERs.  The median [\ion{N}{2}]/\hal\ ratio is larger for the
UV-undetected than the UV-detected LINERs (1.84 vs. 1.29), but the
difference between the two distributions is not highly significant
($P_{KS}=0.16$).  A significant difference is found for
[\ion{O}{3}]/\hbeta: the UV-detected LINERs have median
[\ion{O}{3}]/\hbeta\ = 1.44, while the undetected LINERs have median
[\ion{O}{3}]/\hbeta\ = 1.85, and the KS test yields a 98.6\%
confidence level that the two distributions are different.

A greater contribution from stellar photoionization in the UV-detected
LINERs could be responsible for the lower [\ion{O}{3}]/\hbeta\ ratio
in these objects.  The emission-line data of Ho \etal (1997a) indicate
that the mean [\ion{O}{3}]/\hbeta\ ratio for \ion{H}{2} nuclei is
0.81, while for LINERs in spiral hosts, the mean is 1.89.
Furthermore, the median EW(\hal) for LINERs in spiral hosts is 2.2
\AA, while \ion{H}{2} nuclei have a considerably larger median
EW(\hal) of 17.7 \AA.  A greater contribution of stellar
photoionization in the UV-detected LINERs would decrease
[\ion{O}{3}]/\hbeta, [\ion{O}{1}]/\hal, [\ion{N}{2}]/\hal, and
[\ion{S}{2}]/\hal, while at the same time increasing EW(\hal) and
EW(\hbeta), consistent with our results.  However, because most of the
UV-undetected LINERs have very low EWs of \hbeta\ and [\ion{O}{3}], we
are cautious about interpreting the observed difference in the
[\ion{O}{3}]/\hbeta\ ratio between the UV-detected and undetected
subsamples.  Uncertainties in the starlight subtraction procedure,
which are extremely difficult to quantify, could lead to biases in the
measured ratio of two extremely weak emission lines, and it is
possible that such biases could substantially affect the measured line
ratios.

\subsection{Broad \hal\ Emission}

One prediction of the duty cycle hypothesis of Eracleous \etal (1995)
is that UV emission and broad \hal\ emission should occur
simultaneously.  In this model, the broad-line emission, which
originates within a few light-days of the nucleus, disappears rapidly
if the ionizing continuum shuts off.  A correlation between the
presence of broad \hal\ emission and UV emission would also be
expected in an obscuring torus scenario, at least for those LINERs
which are AGNs, since the UV and broad-line emission would both
originate from regions which are small compared with the size of the
obscuring torus.  Using the broad \hal\ measurements of Ho \etal
(1997b), we have compared the UV and broad-line properties of the
galaxies in our sample.

In the combined WFPC2 + FOC sample, including all galaxies (LINER,
transition, and Seyfert nuclei), the UV-detected galaxies are indeed
more likely to host broad \hal\ emission than are the UV-undetected
galaxies.  Of the UV-detected galaxies (10 total), 4 have broad \hal,
while only 3 of the 28 UV-undetected galaxies, or 11\%, have broad
\hal.  Restricting the sample to LINERs only, broad \hal\ emission
occurs in 3 out of 6 UV-detected galaxies but only in 4 out of 18
UV-undetected galaxies.

This result offers support to models in which UV and broad-line
emission are expected to occur together.  However, the interpretation
is complicated by the fact that some of the detected UV sources are
extended or diffuse regions, in which there is no clear evidence for
an AGN-like UV point source.  If the spatially extended UV sources are
clusters of hot stars, rather than AGNs, then one would not
necessarily expect broad \hal\ emission to be associated with the UV
sources.  Another possibility is that the spatially extended UV
emission may be scattered radiation from an obscured AGN.  This
possibility can be tested with polarimetric observations, and in this
case the UV emission could be accompanied by scattered broad \hal\
emission as well.

The two galaxies in our sample with unresolved nuclear UV sources (NGC
3642 and NGC 4203) both have broad \hal\ emission.  Thus, there may be
an even stronger likelihood for broad \hal\ emission to be found in
nuclei having {\it pointlike} UV emission, but our sample is not large
enough to address this question properly.  Also, because our sample
was selected to exclude galaxies with strong broad \hal\ emission, the
rate of detection of UV emission may be lower in our sample than it
would be in an unbiased sample of LINERs.

\subsection{Optical Morphology}

The most striking feature of the optical images is the nearly
ubiquitous presence of optically thick dust lanes in most of the
sample galaxies.  Particularly intriguing is the fact that the dust
lanes occur in very early-type spirals such as NGC 3607 (S0) and NGC
3166 (S0/a).  The prevalence of dust in the nuclei of early-type
spirals has been noted in WFPC1 imaging surveys (e.g., \cite{phi96}),
but the WFPC2 images show the dust lanes in great detail. In the most
extreme cases, such as NGC 3166 and NGC 3607, the contrast between
adjacent bright and dark regions in the F547M images is as much as a
factor of 2 (i.e., $A_V$=0.75 mag).  Assuming for simplicity that the
geometry of the absorbing dust is a uniform screen, the corresponding
extinction at 2200 \AA\ is a factor of 8.8, or 2.4 mag.  UV sources as
bright as those in NGC 3642 or NGC 4203 would be below our detection
threshold if they were located behind such a large obscuring column.

Detailed surface photometry of the optical images will be presented
elsewhere, but one aspect of the optical images that is relevant for
this study is the ``peakiness'' of the optical nuclei.  In a study of
WFPC1 images of Markarian Seyferts, Nelson \etal (1996) have shown
that the nuclear profiles of the Seyfert 1 galaxies are significantly
sharper and more pointlike than those of the Seyfert 2s.  Their result
adds support to unified models for Seyferts, since the unified schemes
predict that the central source is viewed directly in Seyfert 1s but
blocked by obscuring material in Seyfert 2s.  We have examined the
optical nuclear profiles of the galaxies in our sample, in order to
determine whether such a clear difference exists between the type 1
and type 2 LINERs or between the UV-detected and undetected LINERs.
The two objects in our sample having pointlike UV emission and broad
\hal\ emission, NGC 3642 and NGC 4203, appear to have particularly
sharply peaked optical nuclei, compared to most other galaxies in the
sample, suggesting that the correlation found for Markarian Seyferts
may extend to the LINERs as well.

To quantify the ``peakiness'' of the nuclei, we have used a sharpness
parameter adapted from Nelson \etal (1996), and defined as

\begin{equation}
S = \frac{\sum_i C_i^2}{(\sum_i C_i)^2},
\end{equation}
where $C_i$ is the number of counts in pixel $i$, and the sum is taken
over a square aperture surrounding the nucleus\footnote{Nelson \etal
(1996) measure the sharpness parameter over a circular aperture, while
we have used a square aperture for simplicity.}.  This quantity
measures the contrast in count rates within an aperture without making
any assumptions about the spatial profile of the source.  In the
limiting cases of a source with uniform surface brightness or of a
delta-function source, the sharpness parameter will be $1/N$ (where
$N$ is the total number of pixels within the aperture) or 1,
respectively.  Sharpness measurements were made in apertures of
$3\times3$ and $7\times7$ pixels centered on the nuclei of the
galaxies.  In a few objects (NGC 2655, 3166, 3607, and 5005), it is
unclear whether the nucleus itself is visible, and we have centered
the aperture on the brightest pixel in the nuclear region.  The
sharpness measurements are listed in Table 2.  For comparison, an
artificial PSF has $S(3\times3)=0.19$ and $S(7\times7)=0.11$.  In
contrast to the Markarian Seyfert 1s, many of which have very sharply
peaked nuclei (\cite{nel96}), most galaxies in our sample have
sharpness values only slightly greater than those expected for uniform
distributions.  This result reflects the fact that the central sources
are generally of low luminosity and lie within strong backgrounds of
bulge starlight. 

Despite the fact that the sharpness values are low for all galaxies in
the sample, some systematic trends are apparent.  The median sharpness
of the UV-detected galaxies is greater than that of the undetected
objects (0.1182 vs. 0.1127 in a $3\times3$ pixel aperture), and the KS
test shows that the difference between the two distributions is
significant at the 93\% confidence level.  Similarly, the objects in
our sample having broad \hal\ emission have sharper nuclei than
galaxies with no broad \hal\ (0.1178 vs. 0.1128 in a $3\times3$ pixel
aperture), with a 90\% confidence level for a significant difference
between the two distributions.  For the Markarian Seyferts, Nelson
\etal interpreted their results in terms of the effects of the
obscuring torus; in the type 2 Seyferts, the line of sight to the
nucleus is blocked by the parsec-scale torus, resulting in the diffuse
nuclear morphology.  Among the LINERs, however, the most diffuse
nuclei in our sample are those in which the nuclei are clearly
obscured by 100-pc scale dust lanes, such as NGC 3166 and NGC 3607.
Obscuring tori may occur at smaller scales, but the optical nuclear
sharpness of LINERs may be largely determined by the presence or
absence of large-scale dust lanes covering the nuclei.

\section{Discussion}

One of the key questions we wish to address is whether any of the
LINERs in our sample may have nonstellar, AGN-like UV continua.  The
discovery of compact UV sources in NGC 3642 and NGC 4203, in addition
to the broad \hal\ line observed in these galaxies, suggests that
these objects may be true low-luminosity AGNs.  If we speculate that
the compact UV sources in these two galaxies may have nonstellar,
power-law spectra, are their extrapolated ionizing luminosities
sufficient to power the observed LINER emission?  Following the
analysis of Maoz \etal (1995), we calculate the ratio \nion/\nhal,
where \nion\ is the number of ionizing photons extrapolated from the
2200 \AA\ flux, assuming an $f_\nu \propto \nu^{-1}$ continuum, and
\nhal\ is the number of \hal\ photons emitted.  For Case B
recombination, \nion/\nhal = 2.2 if all ionizing photons are absorbed
in the narrow-line region (NLR) (\cite{ost89}).  Using the
extinction-corrected \hal\ fluxes measured by Ho \etal (1997a), we
find \nion/\nhal = 2.4 and 3.2 for NGC 3642 and NGC 4203,
respectively, indicating a surplus of ionizing photons with respect to
the minimum number required to produce the observed \hal\ emission.
Such a comparison is complicated by the uncertainty in the extinction
toward the central source as well as by calibration uncertainties in
the optical spectrophotometry (e.g., slit losses; see \cite{hfs3} for
a discussion), but the fact that \nion/\nhal$ > 2.2$ in both cases
demonstrates that nonstellar photoionization is a plausible mechanism
for the excitation of the NLRs of these galaxies.  The most important
caveat to this conclusion is that UV spectroscopy is needed to test
whether the UV continua are indeed nonstellar.  It is quite possible
that these sources may be young star clusters; \hst imaging has shown
that many nearby galaxies contain near-nuclear star clusters with
sizes of only a few pc and UV luminosities comparable to those of the
nuclear sources in NGC 3642 and NGC 4203 (\cite{meu95}; \cite{mao96}).

The energy budgets of NGC 6500 and NGC 4569 have been discussed by
Barth \etal (1997) and Maoz \etal (1998).  Massive stars alone cannot
provide the power required to ionize the NLR in NGC 6500, and an
additional ionization mechanism must be present, presumably shocks or
an obscured AGN.  In NGC 4569, the young stellar population appears to
be just sufficient to power the emission lines, provided that very
massive stars are still present in the cluster (\cite{mao98}).  No UV
spectral information is available for NGC 5005 or NGC 7743.  However,
if we speculate that the LINER NGC 5005 may have a UV spectral shape
similar to that of NGC 4569 and the other LINERs studied by Maoz \etal
(1998), then it suffers from a severe deficit of ionizing photons.
Its nuclear UV source is too weak by more than an order of magnitude
to power the NLR luminosity, and another energy source must be present
in addition to the hot stars that presumably provide the observed UV
continuum.  If we assume that the UV continuum in the Seyfert 2 galaxy
NGC 7743 is also produced by an extremely young starburst, then after
correcting for the very substantial internal reddening (\hal/\hbeta =
5.85), the UV continuum would be energetically sufficient to power the
narrow-line emission.  However, it is unlikely that hot stars alone
could produce the higher-ionization Seyfert 2 narrow-line spectrum,
and scattered AGN UV continuum radiation, which has been detected in
other Seyfert 2 nuclei (e.g., \cite{pdr93}), could be the source of
the spatially extended UV emission.

If the UV-undetected galaxies are simply obscured versions of the
UV-detected galaxies, then some of them could also be powered by
nonstellar photoionization.  Observations at other wavebands will be
necessary to determine the nature of these objects: for example, hard
X-ray emission or compact radio sources can reveal the presence of a
hidden AGN.

The possible role of shock excitation within LINER nuclei remains
unclear.  The simulations of Dopita \& Sutherland (1995, 1996) have
shown that fast shocks can closely reproduce the optical spectra of
LINERs, but UV emission-line spectroscopy is a more definitive test of
the shock hypothesis.  At present, the only LINER known to have UV
emission-line ratios consistent with the shock models is the nuclear
disk of M87 (\cite{dop96}), and UV spectroscopy of more objects is
clearly needed to determine whether other LINERs may be excited by
shocks.  Because shocks produce little UV continuum emission,
obtaining UV spectra of {\it UV-undetected} LINERs to search for
high-excitation emission lines is perhaps the best way to assess the
importance of shock heating in LINERs.  Such a test is possible with
the Space Telescope Imaging Spectrograph (STIS).  Morse, Raymond, \&
Wilson (1996) have also pointed out that the \hst F220W filter
passband contains several UV emission lines which are expected to be
strong in fast shocks (as in the models of \cite{ds96}), and which
might contribute to a spatially extended component of the UV emission
detected in \hst\ images; again, UV spectroscopy of more LINERs is
needed to test this possibility.

The observed inclination dependence of the UV detection rate echoes
similar results that have been obtained for Seyfert nuclei.  Keel
(1980) found that Seyfert 1 nuclei are preferentially found in face-on
hosts.  This result was extended by Kirhakos \& Steiner (1990), McLeod
\& Rieke (1995), and Maiolino \& Rieke (1995), who demonstrated that
Seyfert 2 nuclei avoid edge-on hosts as well, and by Simcoe \etal
(1997), who have shown that the distribution of soft X-ray selected
Seyferts is biased against edge-on hosts.  The consensus of these
studies is that the NLR is typically surrounded by optically thick
obscuring material, which must be preferentially aligned with the
galactic plane in a geometrically thick torus-like structure denoted
the ``outer torus'' by Maiolino \& Rieke (not to be confused with the
inner, parsec-scale torus surrounding the broad-line region).  This
outer, 100-pc scale torus of obscuring material prevents the optical
or UV detection of Seyfert nuclei in edge-on or nearly edge-on hosts.
The question of whether LINERs are preferentially found in face-on
hosts is beyond the scope of this paper, but the fact that UV sources
are not detected in edge-on LINERs suggests that the thick outer torus
scenario proposed for Seyferts may apply to LINERs as well.  If we
take 65\arcdeg\ as the maximum inclination at which UV sources may be
detected, then the thickness-to-radius ratio of the outer torus is
roughly unity.  The obscuration need not be in any kind of organized
structure such as a torus, however.  The optical images suggest that
the morphology of the obscuring clouds is in most cases far more
irregular, patchy, or filamentary than a simple torus model.  Two
galaxies in our sample, NGC 3607 and NGC 7217, do have well-defined
dust rings, although it is not clear whether the vertical structure of
these rings is thick or flattened.

Infrared (IR) observations could be used to quantify the dust content
of the nuclei, but sufficient IR data do not yet exist to study the
entire sample.  The spatial resolution of {\it IRAS} measurements is
too coarse to isolate the nuclei on the relevant scales of 100 pc or
less.  It would be interesting to compare the optical dust-lane
properties with ground-based small-aperture mid-IR measurements, but
such detections have been reported for fewer than half of the galaxies
in this sample (see Giuricin \etal 1994 for a compilation of 10 $\mu$m
measurements of nearby galactic nuclei).

Since our results imply that the majority of LINERs contain nuclear UV
sources, most of which are obscured by dust, there is no need to
invoke a low AGN duty cycle, as in the hypothesis of Eracleous \etal
(1995), to explain the low UV detection rate.  However, even if the
rate of stellar breakup flashes is not the explanation for the UV
detection rate, the recent appearance of broad double-peaked \hal\
emission in the LINERs NGC 1097 (\cite{sbw93}), Pictor A
(\cite{he94}), and M81 (\cite{bow96}) serves as an important reminder
that such accretion events may contribute significantly to the
energetics of LINERs.

\section{Conclusions}

Our discovery that the probability of detection of UV emission in a
LINER is dependent on the host galaxy inclination and internal
reddening provides a straightforward explanation for the low UV
detection rate of LINERs.  Nearly all of the UV-detected LINERs are in
low-inclination, low-extinction hosts, while {\it no} LINERs in highly
inclined ($i>65\arcdeg$) or highly reddened (\hal/\hbeta$>3.6$) hosts
are detected in the UV.  The optical images show that many
low-inclination galaxies (such as NGC 3607) have optically thick dust
lanes covering the nuclei which could easily obscure a UV source;
thus, the fraction of LINERs that contain nuclear UV sources is likely
to be greater than the observed fraction, even among low-inclination
hosts.  Since two-thirds of LINERs with $i<45\arcdeg$ are detected in
the UV, it seems reasonable to take this ratio as a lower limit to the
detection rate we would obtain in the absence of any obscuration.  The
dependence of UV detection rate on emission-line equivalent width is
also easily understood if the obscuring material blocking our view of
the UV-undetected nuclei also obscures part of the NLRs in these
galaxies, and if the emission-line luminosity of LINERs is correlated
with their UV luminosity.

If the pointlike UV sources in NGC 3642 and NGC 4203 are AGNs, their
extrapolated continua are energetically sufficient to account for the
emission-line luminosities of the NLRs of these galaxies.  This result
supports the hypothesis that at least some LINERs may be primarily
photoionized objects which are low-luminosity examples of the AGN
phenomenon.

Spectroscopy is needed to determine the origin of the UV emission,
which may arise from low-luminosity AGNs or compact young star
clusters.  The model of Eracleous \etal (1995), in which {\it all\/}
the differences between UV-detected and undetected LINERs can be
ascribed to AGN duty cycles, does not account for the observed
dependence of detection rate on inclination, or for the fact that some
UV sources in LINERs are likely to be young star clusters.  Although
duty cycles may occur in those LINERs which are AGNs, dust obscuration
is likely to be the dominant factor determining whether a UV source is
visible.

\acknowledgements

We are grateful to Dan Maoz for supplying data on the FOC sample of
LINERs, and to Daniela Calzetti for providing template galaxy spectra.
Mike Liu provided valuable assistance and advice on image processing.
Support for this work was provided by NASA grant NAG 5-3556, and by
NASA grant GO-5419.02-93A from the Space Telescope Science Institute,
which is operated by AURA, Inc., under NASA contract NAS 5-26555.
L. C. H. acknowledges fellowship support from the Harvard-Smithsonian
Center for Astrophysics.  This research has made use of the NASA/IPAC
Extragalactic Database (NED), which is operated by the Jet Propulsion
Laboratory, California Institute of Technology, under contract with
NASA.

\begin{center}
Appendix A. Notes on Individual Galaxies
\end{center}

{\it NGC 2655--} Dust lanes cross the nuclear region of this Seyfert 2
galaxy.  The true nucleus may be heavily obscured.

{\it NGC 3607--} The F547M image clearly shows the circumnuclear dust
ring described by Singh \etal (1994), as well as a thick filament of
dust crossing over the nucleus itself.  The inner radius of the dust
ring is 3\arcsec, or $\sim300$ pc. Singh \etal conjecture that the
nuclear dust was accreted during a recent interaction with the
nearby companion NGC 3608.
             
{\it NGC 3642--} Compact UV emission and broad \hal\ emission make
this galaxy an excellent example of an AGN-like LINER.  A compact
nuclear X-ray source with a power-law spectrum was detected by
Koratkar \etal (1995).
             
{\it NGC 3718--} This LINER has the highest nuclear sharpness of any
galaxy in our sample, and is a source of broad \hal\ emission, but no
UV emission was detected.

{\it NGC 4111--} This S0 galaxy is nearly edge-on, with $i=87\arcdeg$.
Surrounding the bright, compact nucleus is a dust lane oriented
perpendicular to the galaxy's plane; its morphology is suggestive of a
polar ring having a radius of 3\arcsec.
             
{\it NGC 4192--} This transition-type spiral is highly inclined
($i=83\arcdeg$) and is heavily reddened internally (\hal/\hbeta\ =
14.3 after removal of Galactic reddening), yet its optical nucleus is
one of the sharpest in the sample.
             
{\it NGC 4203--} Like NGC 3642, this galaxy hosts a compact nuclear UV 
source and broad \hal\ emission.  

{\it NGC 4569--} Maoz \etal (1995) found the UV source in NGC 4569 to
be pointlike, whereas in our image it is clearly extended even at the
smallest radii, as illustrated in Figure 4.  The most likely
explanation for this discrepancy is that the FOC image of NGC 4569 was
taken prior to the first \hst refurbishment mission, and hence
suffered from the aberrated \hst PSF; it was also very heavily
saturated, making profile measurements at small radii extremely
difficult.  Spacecraft jitter cannot be the cause of the spatial
extent of the WFPC2 image, as the spatial profile is identical in the
two UV exposures.  The maximum possible contribution from a single
point source in the WFPC2 image is 23\% of the total flux, but the
fluxes measured in the WFPC2 and FOC images differ by only 10\%. The
F547M image of NGC 4569 is saturated, and we are unable to measure the
optical nuclear sharpness parameter for this object.  The UV source
has $S(7\times7)=0.0489$, compared with $S(7\times7)=0.1531$ for the
synthetic F218W PSF, another indication that the UV profile of NGC
4569 is significantly more extended than a point source.  The {\it
IUE} and optical study by Keel (1996) demonstrated that the nucleus of
this galaxy is a young star cluster 6 magnitudes more luminous than
the core of 30 Doradus, and that the optical light from the cluster is
dominated by A-type supergiants.  \hst data presented by Maoz \etal
(1998) show that the UV spectrum is nearly an exact match to that of
the young starburst cluster NGC 1741-B, which has a likely age of 4--5
Myr (\cite{clv96}).

{\it NGC 5005--} The optical morphology of the nucleus is suggestive
of a clumpy, dusty starburst region.  The faint patch of UV emission
appears at the same location as the peak of the optical emission.
             
{\it NGC 5195--}  This galaxy has been classified both as an amorphous
irregular (I0) and as type SB.  The LINER classification is uncertain,
as the [\ion{O}{1}] and [\ion{O}{3}] lines are extremely weak.

{\it NGC 6500--} The nucleus of this LINER shows kinematic evidence
for a nuclear outflow or wind; see Gonz\'alez Delgado \& P\'erez
(1996) for a discussion of the optical emission lines and kinematics.
\hst Faint Object Spectrograph data show that the UV continuum shape
is consistent with that of a young stellar population with age
$\lesssim100$ Myr, and spectral features due to a young starburst may
be weakly present (\cite{bar97}; \cite{mao98}).  The UV emission-line
spectrum is of low excitation; \ion{C}{3}] $\lambda1909$, \ion{C}{2}]
$\lambda2326$, and \ion{Mg}{2} $\lambda2800$ are the strongest
collisionally excited lines in the spectrum, and \ion{C}{4}
$\lambda1549$ is not detected.  At most $\sim7$\% of the UV emission
from the nuclear region could be due to a single unresolved point
source.  

{\it NGC 6951--} This low-luminosity Seyfert 2 galaxy hosts a
spectacular circumnuclear star-forming ring with projected dimensions
of $1.0 \times 0.7$ kpc$^2$.  The structure of the ring and its
compact star clusters are discussed by Barth \etal (1995).
             
{\it NGC 7217--} A dust ring of inner radius 6\farcs5, or $\sim500$
pc, surrounds a nuclear region which appears to be nearly devoid of
dust.

\bigskip
\bigskip

{\it Note:} Because of the large size of Figures 1 and 2, they are not
included with the astro-ph distribution of this paper.  They can be
found at http://astro.berkeley.edu/$\sim$barth/papers/uv.

{\it Figure 1-- } WFPC2/PC F218W images of the six UV-detected
nuclei.  Image size is 4.6\arcsec$\times$4.6\arcsec.  Logarithmic
stretch was used for NGC 4569, and linear stretch for the other
objects.

{\it Figure 2--} WFPC2/PC F547M images of the galaxies, rotated
so that north is up and east is to the left. Images are
27\arcsec$\times$27\arcsec\ and are displayed with logarithmic
stretch.

\clearpage
\begin{deluxetable}{llrccclcc}
\tablewidth{7in}
\tablenum{1}
\tablecaption{The Galaxy Sample}
\tablehead{\colhead{Galaxy} & \colhead{Type} & \colhead{$cz$} &
\colhead{$i$} & \colhead{$A_V$(Galactic)} & \colhead{H$\alpha$/H$\beta$} & 
\colhead{Nuclear} & \multicolumn{2}{c}{Exp. Time (s)} \nl
\colhead{ } & \colhead{} & \colhead{(km s$^{-1}$)} &
\colhead{(\arcdeg)} & \colhead{(mag)} & \colhead{} & \colhead{Type} &
\colhead{F218W} & \colhead{F547M}
}
\startdata
NGC 1961\dotfill & SABc    & 3934\phn\phn & 54 & 0.29 & 5.44 & L2   & 1800 & 300 \nl
NGC 2655\dotfill & SAB0/a  & 1404\phn\phn & 29 & 0.02 & 4.94 & S2   & 1600 & 300 \nl
NGC 3166\dotfill & SAB0/a  & 1345\phn\phn & 58 & 0.01 & 6.50 & L2   & 1800 & 300 \nl
NGC 3169\dotfill & Sa pec  & 1233\phn\phn & 59 & 0.03 & 5.03 & L2   & 1600 & 300 \nl
NGC 3607\dotfill & S0      & 935\phn\phn  & 30 & 0.00 & 5.56 & L2   & 1600 & 260 \nl
NGC 3642\dotfill & Sbc     & 1588\phn\phn & 38 & 0.00 & 3.21 & L1.9 & 1800 & 300 \nl
NGC 3718\dotfill & SBa pec & 994\phn\phn  & 66 & 0.00 & 4.33 & L1.9 & 1800 & 300 \nl
NGC 3992\dotfill & SBbc    & 1048\phn\phn & 59 & 0.01 & 2.09 & T2:  & 1400 & 230 \nl
NGC 4036\dotfill & S0      & 1397\phn\phn & 72 & 0.02 & 2.86 & L1.9 & 1800 & 300 \nl
NGC 4111\dotfill & S0      & 807\phn\phn  & 87 & 0.00 & 4.78 & L2   & 1800 & 300 \nl
NGC 4192\dotfill & SABab   &$-142$\phn\phn& 83 & 0.11 & 13.02\phn & T2  & 1400 & 230 \nl
NGC 4203\dotfill & SAB0    & 1086\phn\phn & 26 & 0.00 & 1.03 & L1.9 & 1800 & 300 \nl
NGC 4569\dotfill & SABab   &$-235$\phn\phn& 64 & 0.07 & 4.90 & T2   & 1400 & 230 \nl
NGC 4651\dotfill & Sc      & 805\phn\phn  & 59 & 0.02 & 2.95 & L2   & 1800 & 300 \nl
NGC 5005\dotfill & SABbc   & 946\phn\phn  & 64 & 0.00 & 2.59 & L1.9 & 1400 & 230 \nl
NGC 5195\dotfill & SB0 pec & 465\phn\phn  & 46 & 0.00 & 1.90 & L2:  & 1400 & 230 \nl
NGC 6500\dotfill & Sab     & 3003\phn\phn & 46 & 0.28 & 3.31 & L2   & 2200 & 350 \nl
NGC 6951\dotfill & SABbc   & 1426\phn\phn & 28 & 0.66 & 12.41\phn & S2 & 2000 & 300 \nl
NGC 7217\dotfill & Sab     & 946\phn\phn  & 32 & 0.31 & 6.83 & L2   & 1800 & 300 \nl
NGC 7743\dotfill & SB0     & 1710\phn\phn & 38 & 0.13 & 5.85 & S2   & 2000 & 300 
\enddata

\tablecomments{Morphological types and heliocentric redshifts taken
from NED.  Inclinations ($i$) taken from Tully (1988), except for NGC
1961 and NGC 6500.  For these galaxies, the inclination was calculated
from the isophotal diameters given in the RC3 using the method
described by Tully (1988).  Galactic extinction ($A_V$) from Burstein
\& Heiles (1984).  Balmer decrements (H$\alpha$/H$\beta$), corrected
for Galactic reddening, taken from Ho et al. (1997a).  Nuclear
spectroscopic type from Ho \etal (1997a).  L = LINER; T =
LINER/\ion{H}{2} ``transition'' type; and S = Seyfert.  Colon
indicates uncertain classification.  The numerical classification
indicates whether or not broad H$\alpha$ is detected (1.9 = weak
detection; 2 = no detection).}
\end{deluxetable}

\begin{deluxetable}{lrrcc}
\tablewidth{4in}
\tablenum{2}
\tablecaption{Measured Parameters}
\tablehead{
\colhead{Galaxy} & \colhead{$f_{\lambda}$(2200)} &
\colhead{$I_{\lambda}$(2200)} & \colhead{$100 \times S(3)$} & 
\colhead{$100 \times S(7)$} \nl
\colhead{} & \colhead{(1)} & \colhead{(2)} & \colhead{(3)} &
\colhead{(4)}
}

\startdata
NGC 1961\dotfill & $<2$  & $<4$  & 11.29 & 2.20   \nl
NGC 2655\dotfill & $<3$  & $<3$  & 11.19 & 2.18  \nl
NGC 3166\dotfill & $<2$  & $<2$  & 11.12 & 2.05  \nl
NGC 3169\dotfill & $<3$  & $<3$  & 11.27 & 2.28  \nl
NGC 3607\dotfill & $<3$  & $<3$  & 11.15 & 2.07  \nl
NGC 3642\dotfill & $19$  & $19$  & 11.86 & 2.45  \nl
NGC 3718\dotfill & $<2$  & $<2$  & 12.95 & 2.84  \nl
NGC 3992\dotfill & $<3$  & $<3$  & 11.47 & 2.24  \nl
NGC 4036\dotfill & $<2$  & $<2$  & 11.28 & 2.18  \nl
NGC 4111\dotfill & $<2$  & $<2$  & 11.19 & 2.17  \nl
NGC 4192\dotfill & $<3$  & $<4$  & 12.18 & 2.54  \nl
NGC 4203\dotfill & $21$  & $21$  & 12.24 & 2.51  \nl
NGC 4569\dotfill & 1100  & 1300  & \nodata\tablenotemark{a} & \nodata\tablenotemark{a} \nl
NGC 4651\dotfill & $<2$  & $<2$  & 11.64 & 2.29  \nl
NGC 5005\dotfill & $6$   & $6$   & 11.78 & 2.22  \nl
NGC 5195\dotfill & $<3$  & $<3$  & 11.18 & 2.22  \nl
NGC 6500\dotfill & $27$  & $57$  & 11.18 & 2.08  \nl
NGC 6951\dotfill & $<2$  & $<11$ & 11.48 & 2.31  \nl
NGC 7217\dotfill & $<3$  & $<5$  & 11.27 & 2.11  \nl
NGC 7743\dotfill & $9$   & $12$  & 11.62 & 2.38  
\enddata
\tablecomments{Columns (1) Measured flux at 2200 \AA, corrected for
WFPC2 internal throughput degradation, in units of $10^{-17}$ erg
s$^{-1}$ cm$^{-2}$ \AA$^{-1}$.  (2) Flux corrected for Galactic
extinction, in units of $10^{-17}$ erg s$^{-1}$ cm$^{-2}$ \AA$^{-1}$.
(3) $100 \times$ sharpness parameter measured in a $3\times3$ pixel
aperture.  (4) $100 \times$ sharpness parameter measured in a
$7\times7$ pixel aperture.  } \tablenotetext{a}{F547M image
saturated.}
\end{deluxetable}

\begin{deluxetable}{lccl}
\tablewidth{5.5in}
\hoffset+1in
\tablenum{3}

\tablecaption{Emission-Line Properties of UV-Detected and UV-Undetected LINERs}

\tablehead{\colhead{Quantity} & \colhead{Median Value for} &
\colhead{Median Value for} &
\colhead{$P_{KS}$} \nl
\colhead{ } & \colhead{UV-Detected LINERs} & \colhead{UV-Detected LINERs} &
\colhead{ }
}

\startdata
H$\alpha$/H$\beta$\dotfill                     & 3.18 & 4.33 & 0.011 \nl
[\ion{O}{3}] $\lambda5007$/H$\beta$\dotfill    & 1.44 & 1.85 & 0.014 \nl
[\ion{O}{1}] $\lambda6300$/H$\alpha$\dotfill   & 0.24 & 0.25 & 0.86  \nl
[\ion{N}{2}] $\lambda6583$/H$\alpha$\dotfill   & 1.29 & 1.84 & 0.16  \nl
[\ion{S}{2}] $\lambda\lambda6716,6731$/H$\alpha$\dotfill                 & 1.22 & 1.37 & 0.76  \nl
                                       &      &      &       \nl
EW(H$\beta$) (\AA)\dotfill                    & 2.41 & 0.62 & 0.019 \nl
EW([\ion{O}{3}]) (\AA)\dotfill                 & 3.36 & 1.23 & 0.039 \nl
EW(H$\alpha$) (\AA)\dotfill                    & 5.66 & 2.25 & 0.14  

\enddata

\tablecomments{Emission-line data adapted from Ho \etal (1997a).}

\end{deluxetable}

\begin{figure}
\figurenum{3}
\plotone{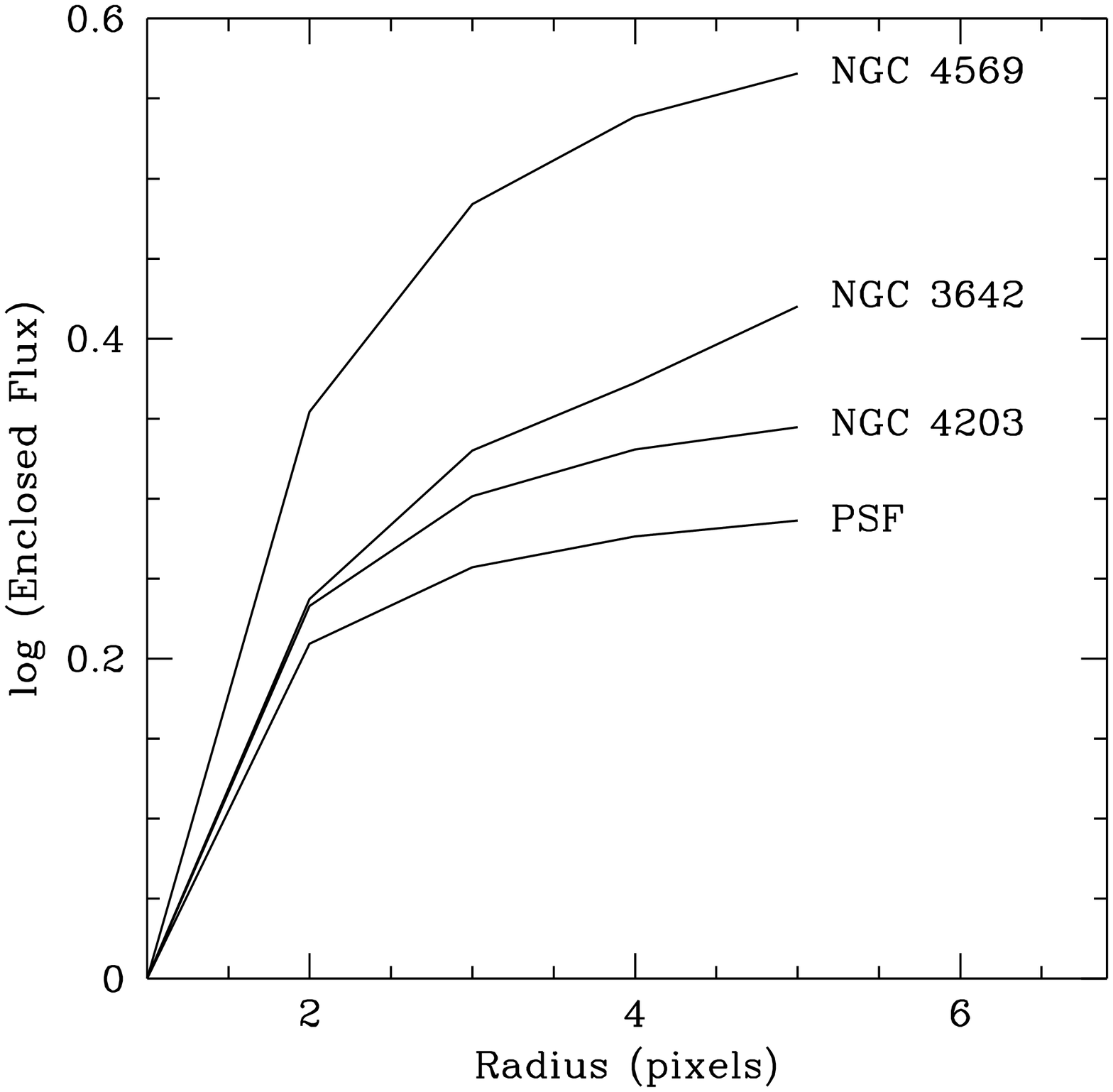}
\caption{Photometric growth curves of the three compact UV sources and
the artificial PSF, normalized to zero at $r=1$ pixel radius.  Note
that NGC 3642 and NGC 4203 have nearly identical profiles within
$r=2$, while NGC 4569 is more extended than the PSF at all radii.}
\end{figure}

\begin{figure}
\figurenum{4}
\plotone{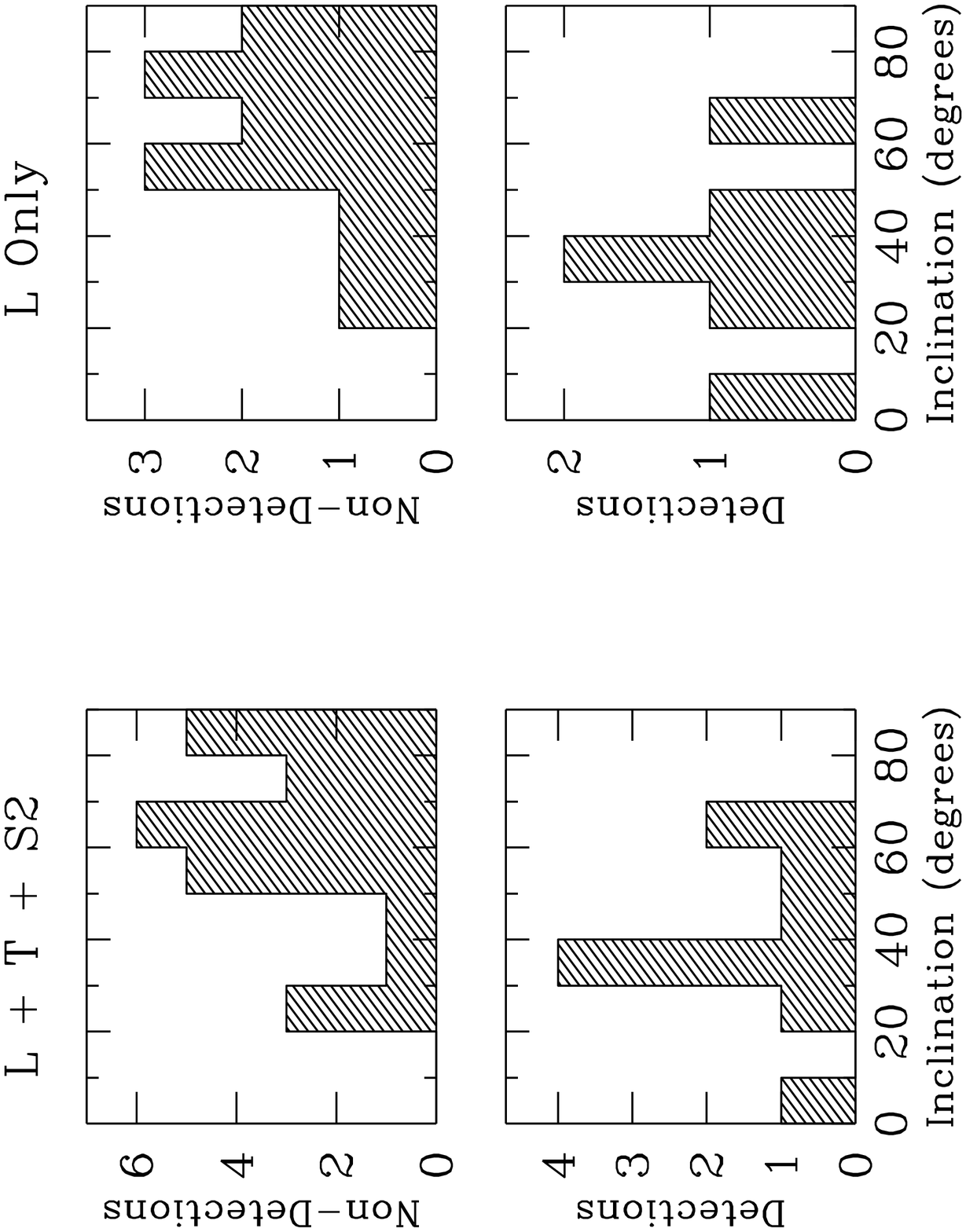}
\caption{Histograms of host galaxy inclination, for spiral and SO
hosts only.  {\it Left panel--} LINERs, Seyferts, and ``transition''
type galaxies combined.  {\it Right panel--} pure LINERs only.}
\end{figure}

\begin{figure}
\figurenum{5}
\plotone{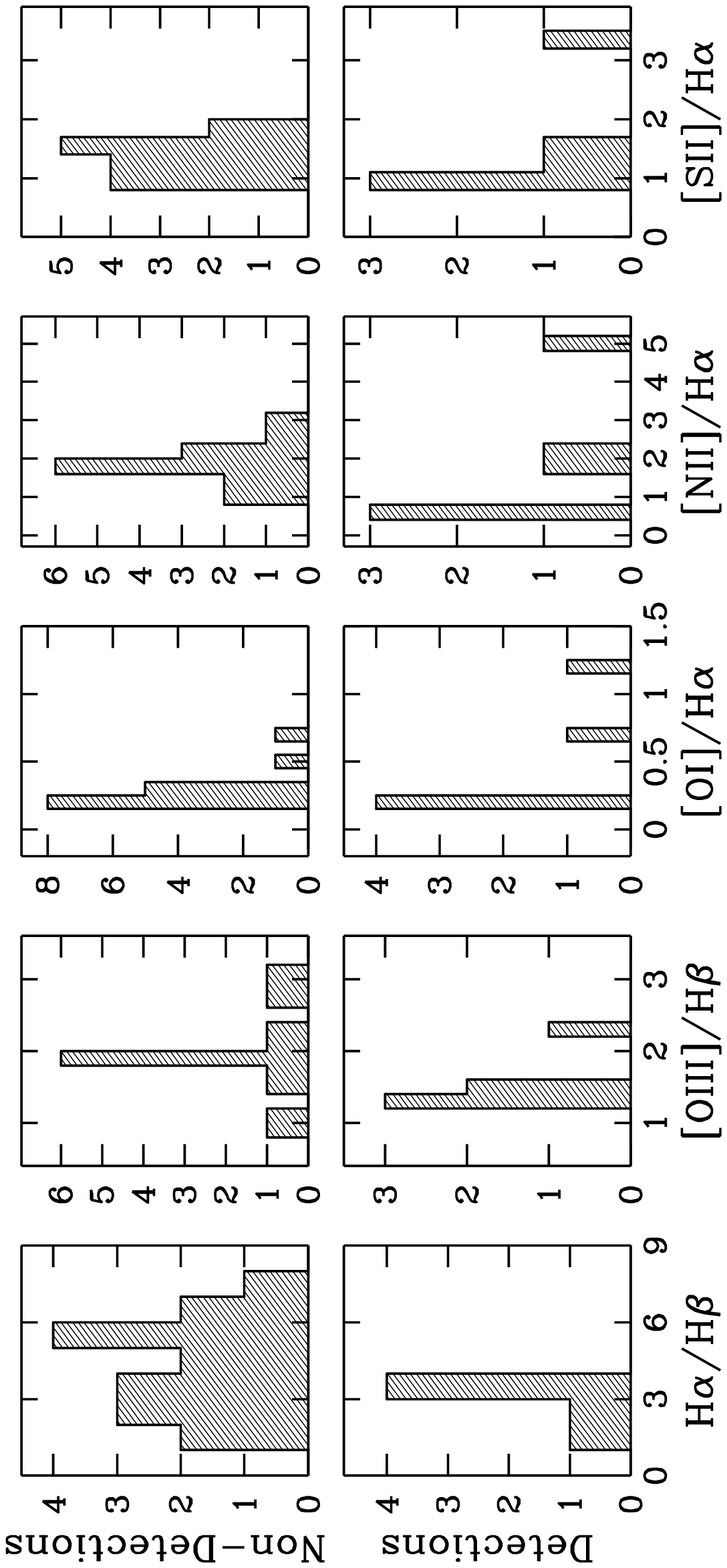}
\caption{Emission-line ratios of UV-detected and undetected LINERs in
the combined WFPC2 + FOC sample.  Data adapted from Ho \etal (1997a).
\hal/\hbeta\ has been corrrected for Galactic extinction, and all
other ratios have been corrected for both Galactic and internal
extinction.}
\end{figure}

\begin{figure}
\figurenum{6}
\plotone{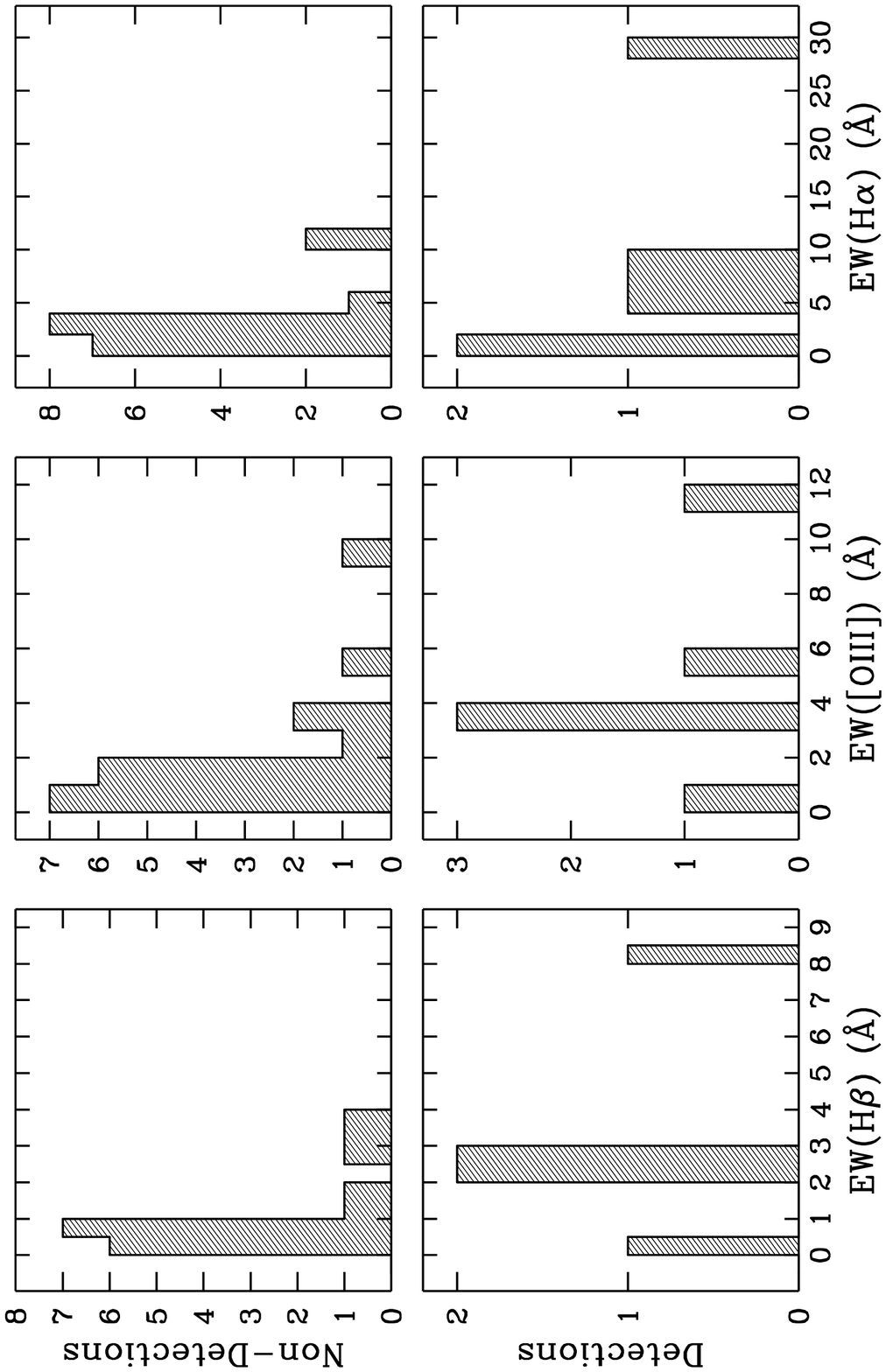}
\caption{Equivalent widths of \hbeta, [\ion{O}{3}]
$\lambda5007$, and \hal, for UV-detected and undetected LINERs in the
combined WFPC2 + FOC sample.  Data adapted from Ho \etal (1997a).}
\end{figure}

\end{document}